\documentclass[sigchi]{acmart}
\usepackage{ bbm, subfig,graphicx} 
\usepackage{algorithm,algorithmic}
\usepackage{amsmath, amssymb, amsthm, graphicx, float}
\usepackage{booktabs}
\usepackage{caption}
\usepackage{multirow}

\AtBeginDocument{%
  \providecommand\BibTeX{{%
    \normalfont B\kern-0.5em{\scshape i\kern-0.25em b}\kern-0.8em\TeX}}}

\copyrightyear{2020}
\acmYear{2020}
\setcopyright{acmcopyright}\acmConference[RecSys '20]{Fourteenth ACM Conference on Recommender Systems}{September 22--26, 2020}{Virtual Event, Brazil}
\acmBooktitle{Fourteenth ACM Conference on Recommender Systems (RecSys '20), September 22--26, 2020, Virtual Event, Brazil}
\acmPrice{15.00}
\acmDOI{10.1145/3383313.3412244}
\acmISBN{978-1-4503-7583-2/20/09}

\begin{document}

\title{Offline Contextual Multi-armed Bandits for Mobile Health Interventions: A Case Study on Emotion Regulation}

%
\author{Mawulolo K. Ameko}
\affiliation{%
  \institution{University of Virginia}
  \city{Charlottesville}
  \country{USA}}
\email{mka9db@virginia.edu}

\author{Miranda L. Beltzer}
\affiliation{%
  \institution{University of Virginia}
  \city{Charlottesville}
  \country{USA}}
\email{beltzer@virginia.edu}

\author{Lihua Cai}
\affiliation{%
  \institution{University of Virginia}
  \city{Charlottesville}
  \country{USA}}
 \email{lc3cp@virginia.edu}

\author{Mehdi Boukhechba}
\affiliation{%
  \institution{University of Virginia}
  \city{Charlottesville}
  \country{USA}}
\email{mob3f@virginia.edu}

\author{Bethany A. Teachman}
\affiliation{%
  \institution{University of Virginia}
  \city{Charlottesville}
  \country{USA}}
\email{bteachman@virginia.edu}

\author{Laura E. Barnes}
\affiliation{%
 \institution{University of Virginia}
 \city{Charlottesville}
 \country{USA}}
\email{lb3dp@virginia.edu}

%
\renewcommand{\shortauthors}{Ameko, et al.}

\begin{abstract}
    Delivering treatment recommendations via pervasive electronic devices such as mobile phones has the potential to be a viable and scalable treatment medium for long-term health behavior management. But active experimentation of treatment options can be time-consuming, expensive and altogether unethical in some cases. There is a growing interest in methodological approaches that allow an experimenter to learn and evaluate the usefulness of a new treatment strategy before deployment. We present the first development of a treatment recommender system for emotion regulation using real-world historical mobile digital data from n = 114 high socially anxious participants to test the usefulness of new emotion regulation strategies.
    We explore a number of offline contextual bandits estimators for learning and propose a general framework for learning algorithms. Our experimentation shows that the proposed doubly robust offline learning algorithms performed significantly better than baseline approaches, suggesting that this type of recommender algorithm could improve emotion regulation. Given that emotion regulation is impaired across many mental illnesses and such a recommender algorithm could be scaled up easily, this approach holds potential to increase access to treatment for many people. We also share some insights that allow us to translate contextual bandit models to this complex real-world data, including which contextual features appear to be most important for predicting emotion regulation strategy effectiveness.
\end{abstract}



\maketitle

\section{Introduction}
Mental illnesses such as depression and social anxiety, if left untreated, can interfere with healthy life functioning, leading to lower disability-adjusted life years~\cite{murray2013state} and higher suicide rates~\cite{Bostwick2000}. 
It is estimated that more than $25\%$ of Americans suffer from a diagnosable mental illness each year~\cite{Kessler2005}, yet half of them  do not receive any treatment~\cite{America} due to the scarce health care resources and limited access to traditional in-person care~\cite{Lin2018}.
New mobile technologies and increasing smartphone ownership give rise to mobile health, a digital health care paradigm that creates opportunities to scale up health interventions to the underserved patient population~\cite{Hilty2013}, especially those with chronic conditions.

One viable target for a digital health intervention that could benefit a significant portion of the population is emotion dysregulation, or difficulty selecting and effectively applying appropriate strategies to modulate the intensity or duration of emotional states~\cite{Gross1998}.
Emotion dysregulation is observed broadly across many mental illnesses, and improvements in emotion regulation~(ER) often accompany decreases in symptom severity ~\cite{Fernandez2016, Sloan2017}. 
The ability to effectively manage negative emotions in our daily lives is of utmost importance. 
For example, days before a job interview, you may not be confident in your preparation, and feel anxious about it. You may find it difficult to focus on anything else, and cannot stop worrying about it or sleep.
To manage your negative emotions, you might try a variety of strategies, including suppressing your thoughts about the upcoming interview, talking to a friend about it, conducting a mock interview for practice, distracting yourself with video games, or taking the advice from your therapist to identify and re-evaluate your catastrophic thoughts. 

Ideally, one would conduct a randomized control trial~(RCT) to evaluate the effect of different ER strategies in different contexts, but this can quickly become unfeasible if the intent is to evaluate more than a dozen strategies across different contexts. We address this challenge in part by using an offline contextual bandits to learn and evaluate a novel treatment recommender algorithm using an observational dataset collected from a population of socially anxious individuals.

While an observational design necessarily limits what causal inferences are possible, our contributions in this work include the following: 1) to the best of our knowledge, this is the first study to apply Contextual multi-armed bandit (CMAB) on ER, a domain that is central to treatment for many mental illnesses; 2) we apply CMAB in an offline setting that learns an interpretable initial policy using observational data; 3) we leverage both passively (e.g., Accelerometer) and actively (e.g., how appropriate was timing of survey) sensed contexts with a designed reward signal using self-reported effectiveness to evaluate the CMAB performance using several different importance sampling based estimators, and compare them with both a random policy and the observed policy. Our results show significantly better performance in the proposed CMAB approaches in terms of the average reward of a policy, which we denote as usefulness.

\section{Related Work}

Emotion regulation (ER) has been studied in psychology for decades due to its importance in understanding how people manage their emotions~\cite{Gross1998}, and its implications for both mental and physical health, and interpersonal relations~\cite{ALDAO2010217}. 
People respond to stressful events using different ER strategies in different social and physical contexts, and according to different situational demands~\cite{Sheppes2014,Dixon-Gordon2015}. 
While ER strategies have long been considered as either adaptive or maladaptive, several researchers have argued that their effectiveness is context dependent ~\cite{Bonanno2013, Aldao2013a}.

Notably, demographic characteristics such as age and gender~\cite{Nolen-Hoeksema2011}, which may be considered internal contexts, significantly influence people's choice of ER strategies. In addition, numerous recent studies have focused on external contexts in people's daily lives, and investigated their impact on ER strategy choice~\cite{Troy2013,Aldao2013,Suri2018}. 
An ecological momentary assessment study by Heiy et al. revealed that many of the most frequently used ER strategies were not the most effective for decreasing negative emotions~\cite{Heiy2014}, suggesting room for improvement in ER even among healthy individuals.
To date, the capability of recommending the most effective ER strategies to people based on different contexts is urgently desired but remains a far-off goal~\cite{Dore2016}.
In this work, we make an effort towards this goal to learn a personalized and adaptive approach for ER strategy recommendation across various contexts.

Many existing works propose various recommender systems targeting different health outcomes. For example, myBehavior, a mobile app that tracks user's physical and dietary habits, recommends personalized suggestions for a healthier lifestyle~\cite{rabbi2015mybehavior}.
Cheung et al.~\cite{cheung2018evaluation} created a mobile app called IntelliCare, which consists of a suite of 12 individual apps as 'treatments' that will be recommended for managing depression and anxiety.
Yang et al.~\cite{yang2018emhealth} created a mobile health recommender system that integrates depression prediction and personalized therapy solutions to patients with emotional distress. In their system, personalization is realized using 9 external factors related to depression, including family life, external competition, interpersonal relationship, self-promotion burden, economic burden, work pressure, individual personality, coping style, and social support, which are assessed using mobile questionnaires.
These mobile health efforts are consistent with a mobile intervention framework called Just-in-time adaptive intervention (JITAI)~\cite{nahum2017just}.

Two aspects regarding the intervention decisions made in a JITAI framework are the timing of intervention delivery and choosing the best intervention strategy to deliver. Most existing works focus on optimizing for the best timing to deliver an intervention (e.g., predicting stressful moments linked to emotional eating~\cite{rahman2016predicting}). By contrast, our work focuses on identifying the most effective ER strategies based on a person's context.
Reinforcement learning with Markov decision processes~(MDPs) are typically used to operationalize the key objectives of a JITAI. Example applications include personalizing sepsis treatment strategies~\cite{peng2018improving}, encouraging physical activity for diabetes patients~\cite{yom2017encouraging}, and managing stress~\cite{jaimes2014stress}. Interestingly, although reinforcement learning is not directly applied to recommend ER strategies for emotion management, it has been applied to understand the psychological and cognitive process of ER~\cite{Marinier2008,Raio2016}.

In this work, we propose to leverage contextual multi-armed bandits, a reinforcement learning algorithm that treats each learning sample as independent from the same underlying data generating the distribution, but ignores the long term impacts on the distal outcome~\cite{dudik2011efficient}. CMAB has been mainly applied in domains such as web contents and advertisement placement~\cite{li2010contextual,tewari2017ads}. In recent years, it has also been applied in numerous mobile health applications, such as hospital and doctor referral for medical diagnosis~\cite{tekin2014discover}, personalized feedback for healthier lifestyle~\cite{rabbi2017towards}, and physical activity recommendation~\cite{liao2020personalized}. Unlike these studies, which were conducted in an online setting or with simulations, our work focuses on the off-policy setting, in which a historical dataset on ER from a mobile health study is used to train an initial warm-start recommendation policy on ER. We design the various reinforcement learning components in the context of recommending ER strategies, and applied various importance sampling based techniques in learning and evaluation.

\begin{figure}
    \centering
    \includegraphics[width=0.45\textwidth]{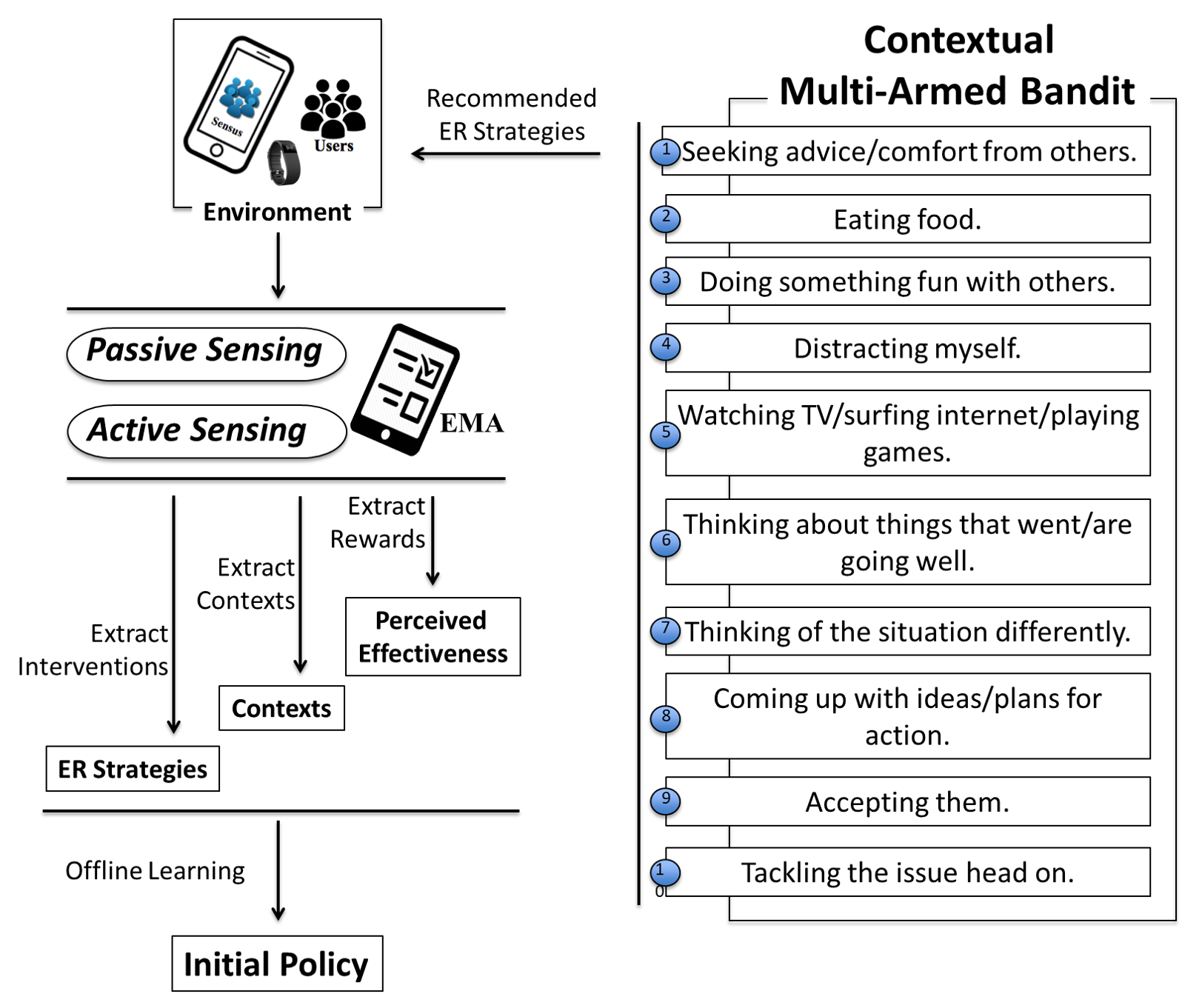}
    \caption{Learning initial policy for emotion regulation (ER) using offline learning in contextual multi-armed bandit.}
    \label{fig:offline_learning_mab}
\end{figure}

\section{Contextual Multi-armed Bandit for Emotion Regulation}

Contextual multi-armed bandit (CMAB) is an reinforcement learning algorithm that leverages contextual information to learn a policy that triggers actions based on the context to achieve optimal expected rewards.
Typically, CMAB consists of an agent that interacts with an environment over a finite number of trials $i = 1,2, \ldots, T$ such that: 1) it observes a context $x$ from an input space $\mathbf{X}$; 2) chooses an action from a set $\mathbf{A} = \{a_1, a_2, \ldots, a_{k-1}, a_{k}\}$, which contains all the strategies that each corresponds to an arm of a k MAB; and 3) receives a reward signal $r_i$.
The goal of the agent is to learn a policy to guide action decisions. 
Unlike a full-blown reinforcement learning algorithm typically modeled using MDPs, where an action decision impacts future states and action selections, CMAB assumes that $\{(x_i, a_i, r_i)\}_{i = 1}^{T}$ are  independently and identically distributed following an unknown generative distribution $\mathbf{D}$.

We formulate ER recommendation as a CMAB using mobile sensing technologies as shown in Figure~\ref{fig:offline_learning_mab}. Smartphones and wearables are applied to track the users both passively with sensor embedded devices and actively with mobile ecological momentary assessments~(EMAs).
These mobile sensing data streams will be processed into the contexts, the recommended ER strategies, and the associated rewards for our CMAB framework.

In the offline learning, logged observational data generated under a different policy will be used to learn and evaluate an initial policy. This data-generating policy is called the behavior policy and can be denoted as $\pi_b$. Similarly, the initial policy is called the target policy denoted as $\pi_e$.

We seek to achieve two objectives: 1) Learn an initial policy $\pi_{e}^*$ given an observational dataset, called the learning problem which is formulated as
\begin{equation}
   \pi_{e}^* =  \operatorname*{argmax}_{\pi_e \in \Pi} \mathbf{V}^{\Pi}.
\end{equation}
Where $\mathbf{V}$ represents the value of a policy and $\Pi$, the function class of possible policies.
2) Evaluate the performance of the initial policy using expected rewards from the testing samples. We call this the evaluation problem and this is formulated as
\begin{equation}
    \mathbf{V}^{\pi_e} = \mathbf{E}_{(x,r) \thicksim D}[r_{\pi_e}(x)].
\end{equation}
In the next section, we present the technical details on both the learning and evaluation problem to learn and evaluate the initial policy.

\section{Learning and Evaluation in Contextual Multi-armed Bandit}

We consider learning in a linear policy class, of which the candidate policies are efficient for learning and easy to interpret. 
We apply importance sampling techniques that use a certain form of weighting scheme denoted as $\frac{\pi_e (a_i|x_i)}{\hat{\pi}_b (a_i|x_i)}$ in context $x_i$ to correct for the distributional shift between the target and behavior policy in order to have an unbiased estimate of the target policy value~\cite{dudik2014doubly}.

There are three main value estimators that lie at the core of offline policy learning and evaluation within the contextual bandit framework; namely, the direct method~(DM), Inverse Propensity Weighting~(IPW), and Doubly-Robust (DR)~\cite{atan2018deep}. None of these approaches are guaranteed to perform optimally in every application scenario. Thus, we apply all of them in learning the optimal policy, and report their results. Below, we provide more details on the benefits and drawbacks of each approach.

\subsubsection*{\textbf{The Direct Method~(DM)}}
The direct method, sometimes called the response surface modeling or covariate adjustment, is the family of approaches that consist of learning a predictive model which maps context and actions to the rewards in a regression model. Specifically, the direct method~(DM) consists of estimating a reward approximator for $\hat{r}(x, a)$, where $\hat{r}: X \times A \rightarrow R$. This will result in a value function:
\begin{equation}
    V_{DM} = \frac{1}{T} \sum \limits_{i=1}^{T}  \pi_e (a_i|x_i) \hat{r}(x_i, a_i) 
\end{equation}\label{dm}
where, $\pi_e$ is the target policy. While this approach is simple to implement and can be used with most regression models, it relies heavily on model specification and overlap in the distributions of the behavior and evaluation policies. This gets even more complex in application domains where the physical process of the underlying environment is not well understood. In effect, most of the direct methods approaches suffer from high bias in the estimates, albeit with low variance for a sufficiently well-specified model.
Some popular examples of algorithms using this approach for learning counterfactual predictions are the Bayesian Additive Regression Trees~(BART)~\cite{chipman2010bart} and the Causal Forest~\cite{wager2018estimation}.

\subsubsection*{\textbf{The Inverse Propensity Weighting Method~(IPW)}}

The inverse propensity weighting approach seeks to correct for the distributional shift caused by the behavior policy by using the behavior policy $\pi_b$ if known or an estimate $\hat{\pi}_b$ (also known as propensity scores~\cite{chakraborty2013statistical}) otherwise. The correction in distribution shifts is achieved using importance sampling in the estimator to evaluate the target policy. Mathematically, a generalized IPW estimator called the trimmed IPW~(tIPW) is as follows:
\begin{equation}\label{ips}
  V_{tIPW} = \frac{1}{T} \sum \limits_{i=1}^{T} \frac{\pi_e (a_i|x_i)}{max\{\hat{\pi}_b (a_i|x_i), \tau\}}  r_i.
\end{equation}

Where $\tau$ is a lower bound on the propensity scores to reduce the effect of large weights on variance of the estimator. When $\tau = 0$ this reduces to a classic IPW estimator. When $\tau = 0$ this approach gives an unbiased estimate of the value of the target policy, however it suffers from high variance due to extreme values of propensity scores (e.g., a propensity score close to zero will give rise to approximately infinite weights). Some examples of algorithms using this approach are the Policy Optimizer for Exponential Models~(POEM)~\cite{swaminathan2015batch} and the Offset tree~\cite{beygelzimer2009offset}.

\subsubsection*{\textbf{The Doubly Robust Estimator~(DR)}}

The doubly robust approach combines the DM and IPW methods to achieve a balanced trade-off between bias and variance. This avoids extremely high bias and variance in the estimator. The DR estimator has been formalized by Dud{\'\i}k et al~\cite{dudik2014doubly} as follows:
\begin{equation}
     V_{DR}  = \frac{1}{T} \sum \limits_{i = 1 }^{T}  \left[ \hat{r}(x_i, a_i) + \frac{\pi_e (a_i|x_i)}{\hat{\pi}_b (a_i|x_i)} \left(r_i - \hat{r}(x_i, a_i)\right) \right].
\end{equation}\label{dr}

The DR estimator combines the DM~(typically a maximum likelihood estimator) with the importance sampling of the residual from the DM approximator. This is described as doubly robust because if the DM model is correct, then the expected residual from the model $E_Y[\hat{\varepsilon}] = 0$, leaving the second term equal to zero for any arbitrary behavior policy $\hat{\pi_b}$; similarly, if the $\hat{\pi_b}$ is correctly estimated, then the second term is a consistent estimator of the error bias from the DM approximator. Though more robust, DR is error prone when both the DM and the behavior policy approximators are misspecified~\cite{kang2007demystifying}.

\begin{algorithm}
 \caption{Generalized Algorithm for Policy Learning}
 \label{policy_learning}
 \begin{algorithmic}[1]
 \renewcommand{\algorithmicrequire}{\textbf{Input:}}
 \renewcommand{\algorithmicensure}{\textbf{Output:}}
 \REQUIRE $\mathbf{X}$, $\mathbf{A}$, $\mathbf{R}$.
 \ENSURE  $\mathbf{\pi^*(x)}$.
 \STATE \textbf{// Propensity Score Estimation}
 \STATE Fit Generalized Boosted Model $\hat{f}: X \rightarrow A$ on $S_{N} = (X, A)$ to balance covariate distribution.
 \STATE Obtain propensity score matrix $\hat{P} = \hat{f}(x)$.
 \STATE \textbf{ // Reward Imputation}
  \STATE fit a one-time logistic regression $\hat{r}: X  \times  A \rightarrow R $ for each strategy
 \FOR { $r_{ij} \in \mathbf{R}$ (a matrix of rewards).}
     \IF {DM method}
         \STATE $\hat{r}^{DM}_{ij} = \hat{r}(x_{ij},a_{ij})$
     \ENDIF
     \IF {IPW method}
         \STATE $\hat{r}^{IPW}_{ij} = \frac{r_{ij}}{\hat{P}_{ij}}$
     \ENDIF
     \IF {DR method}
         \STATE $\hat{r}^{DR}_{ij} = \hat{r}^{DM}_{ij} + \frac{(r_{ij} - \hat{r}_{DM})}{\hat{P}_{ij}}$ 
     \ENDIF
 \ENDFOR
 \STATE Set $\hat{R} = \{ \hat{r}_{ij}\}_{i = 1:T, j = 1:k}$ the weighted reward matrix
 \STATE \textbf{ // Policy Optimization}
 \STATE Fit logistic regression $\hat{h}: x \rightarrow \hat{R}$ on new training set $(X, R)$.
 \STATE \textbf{ // For policy $\pi^*(x)$}
 \STATE $\pi^*(x) = \operatorname*{argmax}_{a \in A} \hat{h}(r_{a}|x)$
 \RETURN $\pi^*(x)$.
 \end{algorithmic} 
 \end{algorithm}

\subsubsection*{\textbf{Propensity Score Estimation}}

As noted above, the behavior policy that generated the data is unknown and needs to be estimated from the data. This is achieved by estimating propensity scores, which represent the likelihood of choosing strategies in different contexts. Propensity scores also serve to reduce multivariate contextual data~\cite{rosenbaum1983central} into one-dimensional scores such that treatment group distributions are matched. The goal of the propensity scores is to create a pseudo-population where the effect of selection bias due to unobserved confounders, as evidenced by distributional mismatch across strategies, is minimized. 

Ensuring overlap in the strategies with respect to the propensity scores reduces the possibility of extreme values in the IPW and DR estimation, given these approaches depend on the estimated score denoted $\hat{\pi}_b (a|x)$ or $\hat{P}_{ij}$ in the algorithm~\ref{policy_learning}. Estimation methods such as logistic regression have typically been used but they are limited due to their linearity assumption~\cite{mccaffrey2013tutorial}. Recently, there are non-parametric machine learning models developed to add more flexibility in order to model more complex data, such as what we usually expect in human data. An example of a non-parametric model is Generalized Boosted Models (GBM). GBM estimation uses an iterative process with multiple regression trees to capture nonlinear relationships between strategies and context variables without over-fitting the data. We implemented GBM propensity score estimation in our analysis using the R package \textit{twang}~\cite{ridgeway2006toolkit}. We used the absolute standardized mean difference~\cite{stuart2013prognostic} as the stopping criteria over 5000 iterations.

\subsubsection*{\textbf{The Learning Algorithms}} In our experiments we used a multivariate logistic regression as the value function approximator that maps contexts to rewards for each ER strategy within the direct method and doubly robust estimators. We used logistic regression with $\ell_2$ regularization for the ease of interpretation and replication in other studies. We will call the learner using direct method (DM) and the one using doubly-robust estimation as (DR) in our experimentation. The offset tree, denoted OT, is different in that it learns several binary regression trees for propensity weighted reward in each offset tree. More details can be found in Beygelzimer et al.~\cite{beygelzimer2009offset}. We compare the performance of these three approaches, and benchmark them against a random policy (i.e., randomly choosing one strategy from the 10 ER strategies) and the observed policy (i.e., what people reported using in the data).

\subsubsection*{\textbf{The Evaluation of Learned Policies}}

Given the selection bias in the test data, we evaluate the performance of the different recommender algorithms using two variants of importance sampling approaches; namely, the inverse propensity weighting~(IPW)(e.i.$\tau = 0$) and the trimmed inverse propensity weighting~(tIPW)(e.i. $\tau \neq 0$) (see equation~\ref{ips}) by varying the parameter $\tau$.

We consider both approaches because while the IPW provides an unbiased estimate of the mean policy reward with possibly high variance, the tIPW reduces the variance at the cost of more bias in the estimator. $\tau$ is a nuisance parameter and can be determined  heuristically if $\tau < 1/k$, where k is the number of strategies according to lemma 3.1 of Strehl et. al,~\cite{strehl2010learning}. We compare the performance of each algorithm on the average reward on the test set.

\subsection{Design of ER Recommender System}
Our contextual variables capture the user's state around the time to use a strategy. They are summarized in Table~\ref{tab:context_var}. A combination of these variables allows us to provide contextual recommendation for ER strategies. For example, given that a user is at home in the evening with a trait social anxiety level of 30, we would recommend tackling issues head on if our algorithm predicts it to be the most effective strategy. 
The actions in our formulation are the top 10 most frequently used adaptive strategies, which are shown in the CMAB in Figure~\ref{fig:offline_learning_mab}. Admittedly, there are multiple ways to reduce dimensionality of the ER feature space and we explored additional approaches in other analyses. However, we chose to focus on this subset of strategies as they are mostly considered healthy strategies (i.e., they tend to be associated with positive health consequences, unlike a strategy such as using alcohol or drugs to change one's feelings) and were most frequently reported in our learning data.

The reward signal needs to reflect the effectiveness of the chosen strategy in the given context at helping to manage the participant's emotion. In our data, participants reported the perceived effectiveness of their ER attempt on a scale of $0$-$10$. We binarized this outcome measure to define a reward signal for the agent. Our threshold was defined as the average of effectiveness scores across all users, or the grand mean. Let $O(x_{i},a_{i})$ denote the immediate effectiveness of the chosen ER strategy at time $i$ in context $x_i$, we have the grand mean as 
\begin{equation}
    \hat{O} = \frac{1}{N} \sum \limits_{i = 1}^{T} O(x_{i},a_{i}) ,
\end{equation}.
The reward signal for each context x and action a is thus defined as: 
\begin{equation}
    r(x,a) = \mathbf{1}_{\{O(x,a) > \hat{O}\}},
\end{equation}
 where $\mathbf{1}$ is an indicator function that returns $1$ when the condition is satisfied, and $0$ otherwise.

\begin{table}[h]
\begin{center}
\caption{Contexts for the proposed contextual multi-armed bandit algorithm.}
\label{table:metrics}
\begin{tabular}{ p{0.3\columnwidth} p{0.6\columnwidth}   } 
 \toprule
Context & Description\\ 
\midrule
Social partners  & self-reported social relationship with people in the context (e.g., being with classmates, friends, strangers/acquaintances, romantic partner and family). \\
\midrule
Social interaction  & self-reported social context (e.g., being alone, no interactions with others or being around them, and interaction with others). \\
\midrule
Social preference &  self-reported social preference (e.g., more people, less people). \\
\midrule
Motivation to change & self-reported motivation to change feelings on a 0-10 scale.  \\
\midrule
Device OS &  device platform~(e.g., Android and iOS). \\
\midrule
Social anxiety score &  self-reported social anxiety score using SIAS scale with 0-80 range.\\
\midrule
Time of day & this is a manual binning of periods of time in the day, (e.g., morning, mid-day, afternoon, and night).\\
\midrule
Semnatic Location & Self-reported locations (e.g., the gym, home, in transition between locations, other homes, other locations, religious places, restaurant, school, shopping center or workplace).  \\
\midrule
Accelerometer &  passively sensed measure of user movement (e.g., mean, energy and standard deviation).\\
\midrule
Activity Type &  passively recognized human activity types (e.g., cycling, stationary, walking and automotive).\\
\midrule
Appropriateness of Timing & self-reported measure of how appropriate the timing was for sending an survey prompt on a scale of 0-10.\\
\bottomrule
\end{tabular}
\label{tab:context_var}
\end{center}
\end{table}

\section{Experiments}

\subsection{Study Design}

After getting approval from the university's Institutional Review Board (IRB), $N = 114$ participants aged 18 years and older were recruited in a US college department and community to enroll in the present study. Participants were eligible to enroll if they scored at least 29 on the Social Interaction Anxiety Scale (SIAS) developed by Mattick \& Clarke~\cite{mattick1998development}. This cutoff was selected to recruit a sample experiencing moderate to severe social anxiety symptoms (scale range is 0-80).
Four participants were excluded in the analysis due to missing data; specifically, 1 participant did not report any EMA data and 3 participants did not have any reports of effectiveness of an ER strategy, leaving $110$ participants with the following demographics: 81 female, 29 male (no participants reported a non-binary gender identity); 86 undergraduates, 11 graduates or professional students, and 13 others; aged 18-34 with mean $20.41$ and SD $2.98$; 82 reported their race/ethnicity as White, 21 Asian, 7 African American, 3 Middle Eastern, 3 Native Hawaiian/Pacific Islander (numbers add up to more than 110 because some participants identified as multiple races).
Their SIAS scores ranged from $29$ to $73$ ($M = 46.68, SD = 10.39$). 
Although the full SIAS was used for recruitment (for comparison to the reference group from prior published work), the sum of the straightforwardly-worded items was used for analyses, because the straightforwardly-worded items have been shown to have preferable psychometric properties to the full scale~\cite{rodebaugh2007reverse}.

A mobile app called MetricWire was installed on all participants’ personal smartphones to collect random time survey data for five weeks.
Six identical surveys were sent randomly within each two hour window from 9am to 9pm daily.
Participants were instructed to complete the surveys as promptly as possible upon receiving the notifications. 
If participants had not completed the survey within 30 minutes of the initial notification, the app sent a reminder notification. If not completed after 45 minutes, the survey disappeared. Participants were instructed to answer the survey with reference to when they received the initial survey notification. This instruction might introduce a small degree of recall bias into survey responses, but was included to enhance ecological validity by sampling a wide variety of situations in daily life, including situations in which it would be difficult to respond to a survey immediately~(e.g., when a participant is taking an exam or in the middle of a conversation). 
Sensor data were also passively collected from participants' smartphones to capture their activity levels and GPS location. Table~\ref{tab:context_var} summarizes the contextual features extracted from both survey and passive data.

\subsection{Data Processing}
We used both the random time survey data and the sensor data from the study to obtain the contexts surrounding the reported ER strategy use and its effectiveness. All contextual variables are aligned with random time prompts using two hour windows. For example, accelerometer data within two hours prior to each survey starting time were aggregated to capture the level of activity for each reported ER strategy use.
We transformed the $x, y, z$ dimensions of the accelerometer using the formula $\frac{1}{3}\sqrt{x^2 + y^2 + z^2}$ to obtain an orientation invariant measure for acceleration.
Activity type data consisted of the activities recognized by MetricWire. These activities include stationary, walking, running, automotive, and cycling. The feature associated with each activity type is the sum total of its occurrence in the two hour window.
Semantic locations such as home, transition, religious place, restaurant, school, shopping, someone else's house, work etc., provided by participants in the surveys were included in the context variables. 
Temporal features were created using four time windows: morning~(9-12PM), mid-day(12 -3PM), late-afternoon~(3-6PM), and night~(6-9PM). 
Finally, we included other survey responses, such as rating the convenience of responding to the prompt when fired, and others summarized in Table~\ref{tab:context_var} as context variables.

The original EMA data consists of $12742$ learning samples from all participants. We excluded samples where participants did not report an effectiveness score for using ER strategies, either because they reported that they did not try to change their feelings (which is one option in the menu provided; 7617 samples were excluded for this reason) or because they used a strategy but skipped the survey prompt about effectiveness (239 samples were excluded for this reason). This leaves  4886 learning samples. 
259 samples where important survey responses were missing (specifically, any missingness on reported convenience of responding to the prompt when fired, semantic location, or motivation to change feelings) were further excluded, leaving 4627 samples for analysis. We avoided imputing the 259 samples as these are self-reported ground truth data. On the other hand, we used multiple chained imputation to impute data on the passively sensed accelerometer and activity type data, which have missing rates of $65\%$ and $68\%$, respectively. The MICE R package with classification and regression trees method was used for the imputation.

The remaining 4627 learning samples consisted of instances where participants reported choosing not only one strategy but also combinations of strategies in a menu of 20 strategies available to them in the survey. Our algorithm, however, considers the effect of a single strategy at a time. To accommodate this constraint, we split the samples in which more than one strategy was reported to have been used into multiple independent samples. For example, if a participant used a combination of eating food and distracting themselves in a given context, we treated this case as two separate samples in which a single strategy was used, and assigned the same effectiveness score to both. This allows us to retain all the data in which effectiveness was reported, increasing power, and not cut the common occurrence in which people report using more than one ER strategy, increasing generalizability. While we recognize this may reduce accuracy in parameter estimation as more bias is being introduced with this data augmentation approach because it is possible that the self-reported effectiveness score does not apply to all applied strategies equally, we felt the benefits for data retention and external validity were worth the trade-off. By augmenting the data this way, we obtain 6259 learning samples, including instances where any of the top 10 most adaptive strategies have been used. By contrast, restricting the data sample to instances where only one strategy was reported being used by the participants, we ended up with 2496 samples, which is about 1/3 of the data generated by the augmentation approach (contact the first author to see results for the CMAB analyses using this smaller dataset). 

We used a total of 40 contextual variables summarized in Table~\ref{tab:effect_table1}, consisting of binary variables (e.g., semantic locations, social partner(s) vs. alone, etc.) and continuous variables, including convenience of responding to the prompt when fired, motivation to change, SIAS score~(SIASsf), activity types, and accelerometer features. The continuous variables were scaled to a range between $[0,1]$ to avoid biasing coefficient estimations toward the continuous variables. 

\section{Results}
Our results as summarized in Table~\ref{tab:results} show the mean reward across the different recommender algorithms and baselines. We report the mean reward with standard errors on a 5-fold cross validation~(due to relatively small data), and test for level of significance using an independent samples t-test at $\alpha = 0.05$. The parameter $\tau$ regulates the effect of extremely large weights due to low propensity scores by capping all scores below the chosen value of $\tau$. Note also that $\tau$ uses the same value in both learning and evaluation for each policy. The algorithm learned with doubly robust estimator~(DR) outperforms all its competitors, including the Offset Tree~(OT) and the DM learner. This can be seen from its absolute mean reward and the tight confidence bounds for all values of $\tau$. This implies that the doubly robust method achieves the right trade-off between high variance and high bias, at least relative to the other approaches tested, making it a more more reliable statistical estimator of off-policy performance in our data. We also see that the gap between the Offset tree and the DR get closer as the value of $\tau$ is increased. This is as expected as the classic OT algorithm is heavily dependent on the inverse propensity weighting and thus more affected by high variance. Notice that the parameter values of $\tau$ are set below $0.1$ to match with the theoretical constraint developed in Lemma 3.1 of \cite{strehl2010learning}. Also note that the DM, Random, and Observed policies are affected by the parameter $\tau$ only in the evaluation stage, but they still benefit from less variance in the mean reward estimation on the test set.  

\begin{table}[h]
    \centering
        \caption{Mean reward by policy~(mean $\pm$ std). Superscripts $\dag$ and $*$ respectively represent statistical significant at $\alpha = 0.05$ over random and behavior policy baselines.}
    \label{tab:results}
    \begin{tabular}{l c c c}
    \hline
        Algorithm & IPW($\tau = 0$)  & IPW($\tau = 0.02$) & IPW($\tau = 0.05$) \\ \hline
        DR & $11.48 \pm 2.07^{ \dag *}$ & $10.84 \pm 1.96 ^{ \dag *}$ & $9.11 \pm 1.27$ \\
        DM & $11.08 \pm 2.38$ & $10.46 \pm 2.26$ & $8.65 \pm 1.28$\\
        OT & $8.60 \pm 1.70$ & $8.41 \pm 1.60 $ & $7.81 \pm 1.42$\\
        Observed & $8.25\pm 0.40$  & $8.20 \pm 0.40$  & $7.84 \pm 0.32$\\ 
        Random & $8.24 \pm 0.51$  & $8.20 \pm 0.50$ & $7.91 \pm 0.49$\\ \hline
    \end{tabular}

\end{table}

\begin{table*}[]
\centering
\footnotesize\setlength{\tabcolsep}{1.5pt}
\caption{Coefficients(\textit{rounded to 2 decimal places}) of Contextual Predictors of Strategies (Strats). The strategies are mapped as follows; \textit{Seeking advice/comfort from others}(S1), \textit{Eating food}(S2), \textit{Doing something fun with others}(S3), \textit{Distracting myself}(S4), \textit{TV/internet/gaming}(S5), \textit{Thinking about things that went/are going well}(S6), \textit{Thinking of the situation differently}(S7),  \textit{Coming up with ideas/plans for action}(S8), \textit{Accepting them}(S9) and \textit{Tackling the issue head on}(S10)}
\label{tab:effect_table1}
\begin{tabular}{c|ccccc|cccccccccc}
\cmidrule[\heavyrulewidth]{1-16} 
     \multicolumn{1}{c}{} & \multicolumn{5}{|c|}{\textbf{Social Partners}}   &  \multicolumn{10}{c}{\textbf{Semantic Locations}}\\
\textbf{Strategies} & Classmates & Friend & Strangers & Romantic & Family & Gym   & Home  & Transit & Other Home & Other Loc & Religious Loc & Restaurant & School & Shopping & \multicolumn{1}{c}{Work} \\
\midrule
S1         & \textbf{-0.18}      & 0.03   & \textbf{0.12}          & -0.03    & 0      & \textbf{0.19}  & 0.01  & \textbf{-0.1}    & 0.03       & -0.05        & 0.04      & \textbf{0.06}       & -0.04  & \textbf{-0.13}    &  -0.03     \\
S2         & \textbf{0.1}        & \textbf{0.16}   & 0             & 0.05     & \textbf{0.03}   & 0.05  & 0.03  & \textbf{0.2}     & \textbf{-0.18}      & \textbf{-0.07}        & \textbf{-0.41}           & \textbf{0.11}       & \textbf{0.14}   &  0        & \textbf{0.13}  \\
S3         & \textbf{0.05}       & \textbf{-0.23}  & \textbf{-0.14}         & \textbf{-0.25}    & \textbf{-0.36}  & \textbf{-0.19} & \textbf{0.06}  & 0.03    & 0   & \textbf{0.18}         &  -0.01  & \textbf{0.09}       &  0.01   & -0.01    & \textbf{-0.15} \\
S4         & \textbf{0.08}       & \textbf{0.05}   & \textbf{0.05}          & \textbf{-0.08}    & 0.05   & \textbf{0.08}  & \textbf{-0.05} & 0.02    &  0          &  -0.03        & \textbf{0.01}            & -0.02      & \textbf{0.03}   &  0    & \textbf{-0.05}      \\
S5         & 0.02       & 0      & \textbf{0.24}          & -0.01    & -0.06  & \textbf{-0.44} & \textbf{0.06}  & \textbf{0.06}    & \textbf{-0.04}      & \textbf{-0.17}        & \textbf{0.04}    & \textbf{0.13}       &  0.05   & \textbf{0.18}     & \textbf{0.13}     \\
S6         & -0.01      & \textbf{0.12} & \textbf{0.07}          & -0.03    & -0.01  & \textbf{0.2}   & -0.01 &  0.03    & \textbf{0.11}       & \textbf{0.09}         & \textbf{-0.54}           & \textbf{-0.19}      & \textbf{0.07} & \textbf{0.16}     &  0.06   \\
S7         & \textbf{-0.07}      & \textbf{0.08}   & 0.02          & \textbf{0.05}     & \textbf{0.07}   & \textbf{0.23}  & \textbf{-0.07} & -0.03   & \textbf{-0.19}      & \textbf{0.07}    & \textbf{-0.06}  & \textbf{-0.2}  & \textbf{0.05}   & \textbf{0.19}     &  0.01  \\
S8         & 0.01       & \textbf{0.04}   & \textbf{0.1}           & -0.04    & -0.02  & \textbf{-0.13} & -0.01 & \textbf{-0.04}   & \textbf{-0.06}  & 0.01 & \textbf{0.26}    & \textbf{-0.09} & \textbf{0.06}   & 0.06  & \textbf{-0.07} \\
S9         & \textbf{0.15}       & \textbf{0.14}   & \textbf{0.06}          & \textbf{0.1}      & \textbf{0.07}   & \textbf{-0.08} & \textbf{-0.12} &  0.04    & \textbf{-0.1}  & -0.01 & \textbf{0.39} & 0.01  & \textbf{-0.15}  & \textbf{0.22}     & \textbf{-0.21}  \\
S10        & 0.01       & \textbf{0.08}   & \textbf{0.05}          & -0.04    & \textbf{0.17}   & 0.06  & \textbf{-0.05} & \textbf{0.1}     &  -0.02      & 0.01   & \textbf{-0.14} &  0.02    &  0.02   & 0.02     &  -0.02 \\ \hline
\end{tabular}
\vspace{0.5cm}

\begin{tabular}{c|ccc|cccc|ccccc}
\cmidrule[\heavyrulewidth]{1-13} 
    \multicolumn{1}{c}{} & \multicolumn{3}{|c|}{\textbf{Other EMA}} & \multicolumn{4}{c|}{\textbf{Time of Day}} & \multicolumn{5}{c}{\textbf{Activity Types}}\\
\textbf{Strategies} & Appropriate & Motiv2Change & SIASsf & Morning & Mid-Day & Afternoon & Night & Stationary & Walking & Running & Automotive & Cycling \\
\midrule
S1         & \textbf{1.07}  &  -0.11  &  -0.12  &  -0.01     & \textbf{0.06}    & \textbf{0.03}   & \textbf{-0.08} & \textbf{1.07}       & \textbf{-1.3}    & \textbf{0.35}    & \textbf{0.67}       & \textbf{0.27} \\
S2       &  -0.08  & \textbf{-0.16} & \textbf{-0.37}  &   0   & \textbf{-0.01}   &    \textbf{-0.05}    & \textbf{0.06} & \textbf{0.26}       & \textbf{0.89}    & \textbf{-0.07}   & \textbf{0.62}       &  -0.05 \\
S3         & -0.12  & \textbf{0.85} &  -0.13  &    \textbf{-0.12}   &  0.03    &  \textbf{0.05}  & 0.03 & 0.33       & \textbf{0.81}    & \textbf{-0.09}   & -0.06      & \textbf{-0.07}  \\
S4        & \textbf{0.69} & \textbf{-1.11}  & \textbf{-0.51}  & \textbf{-0.03}     & \textbf{0.04}    &  \textbf{-0.01}  &  0   & \textbf{0.71}       &  0.1    & \textbf{-0.04}   & \textbf{0.6}        & \textbf{-0.05}   \\
S5      &  0.29 &  -0.02 &  -0.09  &   0.01     &  0       &   0.01   & \textbf{-0.01} & \textbf{-0.15}      &  -0.1    &  0       & \textbf{0.37}       &  0.03 \\
S6        & \textbf{0.74}  & \textbf{-0.39} & \textbf{-0.29}  &  \textbf{-0.1} &  0.02    &  0.01    & \textbf{0.07} & \textbf{0.28}       & \textbf{0.23}    & \textbf{0.35}    & \textbf{0.39}       & \textbf{-0.19} \\
S7        & \textbf{1.17} & \textbf{-0.44}   & \textbf{-0.55}  & 0.01     & \textbf{-0.02}   &  \textbf{-0.02}    & \textbf{0.04} & -0.15      & \textbf{-0.74}   &  0       & \textbf{0.53}       & \textbf{-0.07}  \\
S8      &  \textbf{0.7} & \textbf{-0.29} & \textbf{-0.85}  &  \textbf{-0.07} & \textbf{0.01}    &  0  & \textbf{0.06}  & \textbf{0.5}        & 0.01    & \textbf{0.53}    & \textbf{0.52}       & \textbf{-0.49}\\
S9       & \textbf{0.84}  &  0.21    & \textbf{-0.47}  &   \textbf{-0.03}    & \textbf{0.03}    & -0.01   & 0.01 & 0.21       & \textbf{-0.5}    & \textbf{0.64}    & \textbf{0.89}       & \textbf{0.37}  \\
S10        &  \textbf{0.84}        &  -0.12        & \textbf{-0.35}  &  \textbf{0.03}         &  -0.01   & 0    & \textbf{-0.02} & \textbf{-0.61}      & -0.06   & \textbf{-0.24}   & \textbf{0.65}       & \textbf{-0.6} \\
\bottomrule
\end{tabular}
\vspace{0.5cm}

\begin{tabular}{c|ccc|cc|ccc|ccccc}
\cmidrule[\heavyrulewidth]{1-14}
 \multicolumn{1}{c}{} & \multicolumn{3}{|c|}{\textbf{Accelerometer}} & \multicolumn{2}{c|}{\textbf{Platforms}}
& \multicolumn{3}{c|}{\textbf{Social Interactions}}  &  \multicolumn{5}{c}{\textbf{Social Preference}} \\
\textbf{Strategies} & Mean Acc & Std Acc & Energy Acc & Android & iOS & Alone & Interacting & Around & A lot Fewer & Slightly Fewer & Same & More  &  A Bit More  \\ 
\midrule
S1 & 0.29     & \textbf{-1.4}    & \textbf{-0.67} & 0.03 & -0.03 & \textbf{-0.09} & \textbf{0.13}  & \textbf{-0.04}  & \textbf{-0.1}  & \textbf{-0.14} & \textbf{0.08} & \textbf{0.14}   & 0.01 \\ 
S2 & -0.21    & \textbf{-0.84}   & \textbf{-1.3} & \textbf{0.06} & \textbf{-0.06} &  -0.05 &  0.02  & 0.03   & \textbf{-0.05} & \textbf{-0.13} & \textbf{0.13} & \textbf{-0.14} &  \textbf{0.19} \\
S3  &  \textbf{-0.61}    &  0.1     &  -0.11   & \textbf{0.11} & \textbf{-0.11} & \textbf{-0.18} & \textbf{0.1}   & \textbf{0.08}   & -0.07 & \textbf{0.04} & \textbf{0.21} & \textbf{-0.21} &  0.04  \\
S4 &  \textbf{0.84}     & \textbf{-0.64}   &  0.13 & 0.01 & -0.01  &  -0.01 & \textbf{0.08} & \textbf{-0.07}  & \textbf{-0.16} & \textbf{-0.08} & \textbf{0.08} & \textbf{0.09}  &  \textbf{0.08}  \\
S5 & \textbf{0.51}     & \textbf{-1.13}   & \textbf{-0.81} &  \textbf{0.04} & \textbf{-0.04} & \textbf{-0.09} & \textbf{0.13}        & \textbf{-0.04}  & \textbf{-0.17} & \textbf{-0.14} & \textbf{0.06} & \textbf{0.14}  &  \textbf{0.1}   \\
S6 & \textbf{0.59}     & \textbf{0.45}    &  -0.11 & 0.02 & -0.02 & \textbf{0.04}  &  0    & \textbf{-0.04}  & \textbf{0.06}  & -0.04 & \textbf{0.09} &  \textbf{-0.15} & \textbf{0.05}          \\
S7 & \textbf{0.35}     & \textbf{-0.24}   & 0.22 & \textbf{0.02} & \textbf{-0.02} &  \textbf{-0.07} & \textbf{0.09} & -0.01  &  0.01   & \textbf{-0.07} & \textbf{0.1}  & \textbf{-0.15} &  \textbf{0.11}      \\
S8 &  0.18    & \textbf{-0.47}   & \textbf{0.29} & \textbf{0.05} & \textbf{-0.05} &  -0.01 & \textbf{0.04} & \textbf{-0.03}  & \textbf{-0.08} & -0.03 & \textbf{0.11} & \textbf{-0.1}  &  \textbf{0.1}   \\
S9 &  0.07    & \textbf{0.34}    & \textbf{1.45} & 0 & 0 &  \textbf{0.02}  & \textbf{0.02}  & \textbf{-0.03}  & -0.02 & \textbf{-0.06} & \textbf{0.11} & 0.02  & \textbf{-0.04}        \\
S10 & \textbf{-0.89}    &  0.03    & \textbf{-0.68} & -0.01 & 0.01 &  0.01  & \textbf{0.04}  & \textbf{-0.05}  & \textbf{-0.09} & -0.01  & \textbf{0.05} & -0.03 &  \textbf{0.08}        \\
\bottomrule
\end{tabular}
\end{table*}

To probe deeper into a qualitative evaluation of the DR algorithm, we examine the effect sizes of several contextual variables in the learning stage in terms of how they predict individual strategies. These effect sizes are summarized in Table~\ref{tab:effect_table1}. Contextual variables with a positive effect size can be interpreted as increasing the odds of positive rewards if that strategy is chosen within that context and vice versa for negative effect sizes. For example, the chances are high the strategy will be perceived as effective if the user is recommended to seek advice or comfort from others when they have recently been stationary because the effect size is $1.07$. Note that the effect sizes in bold are statistically significant at $\alpha = 0.05$. 

\begin{figure}
    \centering
    \includegraphics[width=0.45\textwidth]{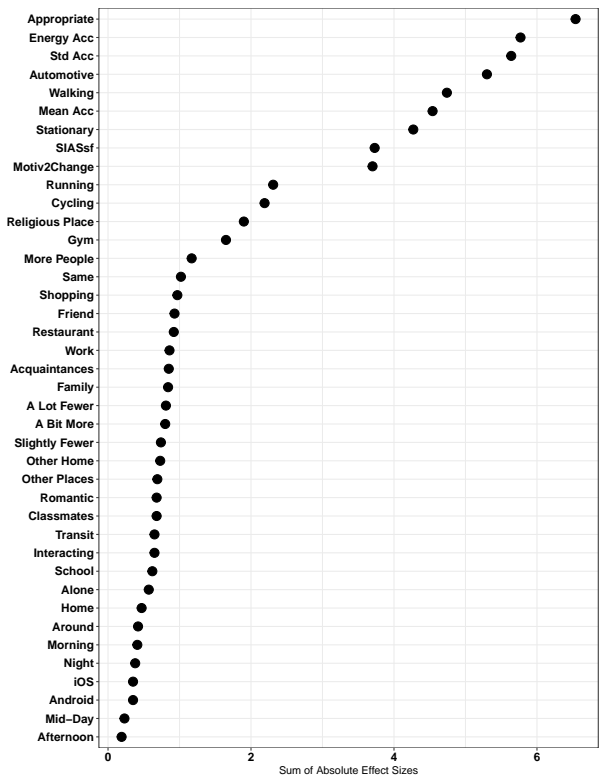}
    \caption{Ranking of contextual variables showing most critical features in determining the effectiveness of strategies. The ranking is based on the sum of absolute values of effect sizes.}
    \label{fig:featureRank}
\end{figure}

While there are many significant effects, pointing to the importance of many contextual factors in ER, a few context variables are notable for their large effect sizes. Overall, the contextual predictors that tended to have the largest absolute effect sizes (indicating that they are the most important in determining effectiveness) are the convenience of responding to the prompt when fired, motivation to change thoughts/feelings, trait social anxiety symptoms, accelerometer features, and certain activity types (see Figure~\ref{fig:featureRank} for a ranking of contextual features from most to least important, as defined by the absolute value of their effect sizes). This suggests that a person's movement helps to determine what ER strategies are most likely to help them feel they have effectively regulated their emotions. The predicted effectiveness of strategies increased when it was a convenient time to be interrupted with a survey prompt, pointing to the importance of timing in interventions (and suggesting that JITAIs may be a step in the right direction). Notably, effect sizes for time of day were smaller than effect sizes for convenient time for interruption, suggesting personalized timing for ER strategy implementation may be particularly helpful. Strategies were predicted to be less effective for more (vs. less) socially anxious participants, even among this sample where all participants were elevated in social anxiety symptoms at baseline), providing further evidence of emotion dysregulation in this population. Higher motivation to change thoughts/feelings predicted higher effectiveness ratings tied to the ER strategy 'doing something fun with others,' but lower effectiveness ratings tied to the ER strategy distraction, demonstrating that contexts can change the effectiveness of different strategies in opposing directions. 

\section{Discussion}
This study provides evidence that a contextual bandits recommender algorithm may be used to improve ER, based on the current finding that the best performing algorithm, the learner with doubly robust estimation~(DR), outperforms the observed ER of socially anxious participants. Further, contexts matter for effective ER, based on our finding that the DR algorithm also outperforms the random algorithm. 

The results from this paper have broad implications for the design and analysis of future recommender systems algorithms. By leveraging the abundance of available observational data from previous studies or interactive systems, a researcher might be able to estimate the usefulness of a novel recommender algorithm before deployment. Recent theoretical studies~\cite{zhang2017transfer} suggest that combining offline policy learning together with online approaches leads to data efficient exploration and adaptations in the online setting. This could potentially reduce the user attrition or disengagement problem that plagues most interactive systems and ecological momentary assessment studies~\cite{tewari2017ads}. In addition, a researcher could use this method to determine the most critical features that affect the effectiveness of ER strategies in order to collect the most salient data for a new study when resources are limited.

Some of the strategies included in this recommender algorithm are cognitive, meaning that they involve a change in thinking (e.g., accepting thoughts/feelings), whereas others are behavioral, meaning that they involve a change in actions (e.g., eating food). Notably, contexts do not seem to have the same effect on strategies of the same cognitive/behavioral type. For example, our findings indicate that walking makes it more likely that thinking about things that went/are going well will be an effective strategy, and less likely that thinking of the situation differently will be effective. The distinction between these two specific strategies is subtle; for one, you are trying to think of positive things that may or may not be related to the situation at hand, and for the other, you are focused on the situation at hand but trying to notice other aspects of it or conceptualize it in a different way. 

Regarding the social strategies in this recommender algorithm (those that use other people to change emotions; e.g., seeking advice/comfort from others), some surprising patterns emerged with social context variables, though with small effect sizes. For example, seeking advice/comfort from others was more likely to help when a user was around strangers and less likely to help when a user was around classmates. While it might be expected that this would be a more helpful strategy when a user was around friends, a romantic partner, or family, none of these contexts had significant effects on this strategy, suggesting that it would be interesting to see whether these patterns would persist if these recommendations were deployed to users. One interesting question that cannot be answered with the current study is how people sought social support; it is possible that people texted or called a friend when they were around strangers, so they may have still used friends to regulate even when those people were not immediately available in their physical environment. Strategies were generally predicted to be more effective when users were interacting with others than when they were alone or around others but not interacting with them, suggesting that the involvement of others might help users regulate effectively.

The effect sizes in Table~\ref{tab:effect_table1} have some implications for the design of future studies. In order to minimize participant burden and to maximize the usefulness of the data collected, a researcher might focus more on collecting the most important features (i.e., those that have the largest effect sizes in predicting the 10 strategies considered in this paper). The most important features were appropriateness of the time to be interrupted; energy, standard deviation, and mean of acceleration; whether participants had recently been in a car, walking, or stationary, their social anxiety symptom severity, and their motivation to change their thoughts/feelings. These important contextual features are all either continuous or discrete variables with many possible values; the less important contextual features are binary. This suggests that future researchers may aim to maximize the predictive value of their contextual variables by considering contextual variables with more variability in their values, as opposed to binary variables. Many of these more important contextual features also reflected movement, so future researchers may wish to preferentially include sensors that capture information about motion. 

The current algorithms work to maximize \textit{short-term }perceived effectiveness of regulating emotions, given the ER strategy attempt and effectiveness rating are reported close in time. However, psychologists have noted that both short-term and long-term regulation are important, with strategies differing in their effectiveness at different timescales \cite{freitas2000regulating}. For example, if you are anxious about an assignment due in a few days, watching TV might make you feel better for 30 minutes but leave you feeling anxious the next day, whereas tackling the issue and starting the assignment might feel worse for the next 30 minutes but make you feel better the next day. While CMAB optimizes for short-term ER effectiveness, evaluating the algorithms for longer-term effectiveness, examining a wider range of ER effectiveness indicators, and examining the algorithms in more diverse samples may all be beneficial directions for future work. Another limitation of this work is that when this policy is deployed, a user will initially need to request an intervention before the most contextually effective strategy is suggested; ultimately, the goal is to be able to passively determine future emotional states and send interventions without the user's initiation.

\section{Conclusion}
In this work, we present a novel application for contextual bandits to learn contextually effective strategies for ER. Our approach is distinct from most existing work in health recommender systems in that we learn an initial policy that might have a positive impact on user engagement when finally deployed, as well as on sample efficiency in the online setting. Our results demonstrate that an experimenter can use available observational data to learn the usefulness of a new intervention policy; this may provide an efficient way to generate hypotheses that can later be tested in (resource intensive)  randomized clinical trials. Given that ER is impaired across many mental illnesses, this work has the potential to enhance the availability of scalable interventions that can be used in daily life for many people.

\begin{acks}
The authors would like to thank the reviewers for their helpful comments. The research presented in this paper was supported by the Data Science Institute of the University of Virginia through their fellowship award, and by the University of Virginia Hobby Postdoctoral and Predoctoral Fellowship Grant and the National Institute of Mental Health R01MH113752 grant.
\end{acks}

\bibliographystyle{ACM-Reference-Format}
\bibliography{references}

\end{document}